\documentclass[conference]{IEEEtran}
\usepackage{cite}

\usepackage{siunitx}

\ifCLASSINFOpdf
\usepackage{graphicx}
 \graphicspath{{../pdf/}{../jpeg/}}
\DeclareGraphicsExtensions{.pdf,.jpeg,.png,.eps}
\else
\fi
\usepackage{amsmath}
\usepackage{amssymb}
\usepackage{mathtools}

\usepackage{algorithmic}
\usepackage{algorithm}
\newsavebox{\ieeealgbox}

\usepackage[tight,footnotesize]{subfigure}
\usepackage[acronym]{glossaries}
\newacronym{mimo}{MIMO}{multiple-input multiple-output}
\newacronym{v2v}{V2V}{vehicle-to-vehicle}
\newacronym{gscm}{GSCM}{geometric-based stochastic channel model}
\newacronym{dmc}{DMC}{diffuse multipath components}
\newacronym{sp}{SP}{specular path}
\newacronym{ekf}{EKF}{extended Kalman filter}
\newacronym{sage}{SAGE}{space-alternating generalized expectation-maximization}
\newacronym{io}{IO}{interaction objects}
\newacronym{mle}{MLE}{maximum likelihood estimator}
\newacronym{blue}{BLUE}{best linear unbiased estimator}
\newacronym{tx}{TX}{transmitter}
\newacronym{rx}{RX}{receiver}
\newacronym{ddtf}{DDTF}{double directional transfer function}
\newacronym{apdp}{APDP}{average power delay profile}
\newacronym{tdoa}{TDoA}{time delay of arrival}
\newacronym{dod}{DoD}{direction of departure}
\newacronym{doa}{DoA}{direction of arrival}
\newacronym{3d}{3D}{3-dimensional}
\newacronym{4d}{4D}{4-dimensional}
\newacronym{cdf}{CDF}{cumulative distribution function}
\newacronym{hrpe}{HRPE}{high resolution parameter estimation}
\newacronym{mpc}{MPC}{multipath component}
\newacronym{fim}{FIM}{Fisher Information Matrix}
\newacronym{em}{EM}{Expectation Maximization}
\newacronym{tdm}{TDM}{time-division multiplex}
\newacronym{snr}{SNR}{signal-to-noise ratio}
\newacronym{pdp}{PDP}{power delay profile}
\newacronym{ofdm}{OFDM}{orthogonal frequency division multiplexing}
\newacronym{lo}{LO}{local oscillator}
\newacronym{pps}{PPS}{pulse per second}
\newacronym{pa}{PA}{power amplifier}
\newacronym{los}{LoS}{line-of-sight}
\newacronym{gps}{GPS}{global positioning system}
\newacronym{siso}{SISO}{single-input single-output}
\newacronym{uca}{UCA}{uniform circular array}
\newacronym{eadf}{EADF}{effective aperture distribution function}
\newacronym{aps}{APS}{angular power spectrum}
\newacronym{t2t}{T2T}{truck-to-truck}
\newacronym{dpr}{dPR}{diffuse power ratio}
\newacronym{ifft}{IFFT}{inverse fourier transform}
\newacronym{cmd}{CMD}{correlation matrix distance}
\newacronym{simo}{SIMO}{single input multiple output}
\newacronym{nlos}{NLoS}{non-line-of-sight}
\newacronym{pl}{PL}{pathloss}
\newacronym{crlb}{CRLB}{Cramer-Rao lower bound}
\newacronym{rmse}{RMSE}{root mean squared error}
\newacronym{uwb}{UWB}{ultra-wideband}
\newacronym{sa}{SA}{simulated annealing}
\newacronym{nsl}{NSL}{normalized sidelobe level}
\newacronym{mse}{MSE}{mean squared error}

\newcommand{\ie}{\emph{i.e.}}
\newcommand{\etal}{\emph{et al. }}

\newcommand{\RNum}[1]{\uppercase\expandafter{\romannumeral #1\relax}}
\newcommand{\tx}[1]{\text{#1}}
\newcommand{\B}[1]{\textbf{#1}}
\newcommand{\BS}[1]{\boldsymbol{#1}}
\usepackage[utf8]{inputenc}
\usepackage[table]{xcolor}
\usepackage{booktabs}
\usepackage{gensymb}
\usepackage{siunitx}

\usepackage{amsthm}
\theoremstyle{definition}

\begin{document}
%
\title{Antenna Switching Sequence Design for Channel Sounding in a Fast Time-varying Channel}

\author{\IEEEauthorblockN{Rui Wang$^1$, \textit{Student Member, IEEE}, Olivier Renaudin$^{1,2}$, \textit{Member, IEEE}, C. Umit Bas$^1$, \textit{Student Member, IEEE}, \\ Seun Sangodoyin$^1$, \textit{Student Member, IEEE}, Andreas F. Molisch$^1$, \textit{Fellow, IEEE}}
\IEEEauthorblockA{$^1$University of Southern California, Los Angeles, CA USA\\$^2$Austrian Institute of Technology, Vienna, Austria}}


%


\maketitle

\begin{abstract}
This paper investigates the impact of array switching patterns on the accuracy of parameter estimation of multipath components for a time division multiplexed (TDM) channel sounder. To measure a fast time-varying channel, the conventional uniform array switching pattern poses a fundamental limit of the number of antennas that a TDM channel sounder can utilize. We propose a method, which is based on simulated annealing, to find non-uniform array switching patterns for realistic antenna arrays, so that we can extend the Doppler estimation range of the channel sounder by suppressing the high sidelobes in the spatio-temporal ambiguity function. Monte Carlo simulations demonstrate that the optimal switching sequence leads to significantly smaller root mean square errors of both direction of departure and Doppler. Results can be applied in both vehicle-to-vehicle and millimeter wave MIMO channel measurements.
\end{abstract}


%
\IEEEpeerreviewmaketitle

\section{Introduction}
Realistic radio channel models are essential for development and improvement of communication transceivers and protocols \cite{molisch2012wireless}. Realistic models in turn rely on accurate channel measurements. Various important wireless systems are operating in fast-varying channels, such as millimeter wave (mmWave), \gls{v2v} and high speed railway systems. The need to capture the fast time variations of such channels creates new challenges for the measurement hardware as well as the signal processing techniques.

Since most modern wireless systems use multiple antennas, channel measurements also have to be done with \gls{mimo} channel sounders. There are three types of implementation of such sounders: (i) full \gls{mimo}, where each antenna element is connected to a different RF chain, (ii) virtual array, where a single antenna is moved mechanically to emulate the presence of multiple antennas, and (iii) switched array, also known as \gls{tdm} sounding, where different physical antenna elements are connected via an electronic switch to a single RF chain. The last configuration is the most popular in particular for outdoor measurements, as it offers the best compromise between cost and measurement duration.

From measurements of MIMO impulse responses or transfer functions, it is possible to obtain the parameters (\gls{dod}, \gls{doa}, delay, and complex amplitude) of the \glspl{mpc} by means of \gls{hrpe} algorithms. Most \gls{hrpe} algorithms are based on the assumption that the duration of one \gls{mimo} snapshot (the measurement of impulse responses from every transmit to every receive antenna element) is shorter than the coherence time of the channel. Equivalently, this means the \gls{mimo} cycle rate, i.e. the inverse of duration between two adjacent MIMO snapshots, should be greater than or equal to half of the maximal absolute Doppler shift, in order to avoid ambiguities in estimating Doppler shifts of \glspl{mpc}. Since in a switched sounder the MIMO snapshot duration increases with the number of antenna elements, there seems to be an inherent conflict between the desire for high accuracy of the DoA and DoD estimates (which demand a larger number of antenna elements) and the admissible maximum Doppler frequency. 

Yin \etal were first to realize that it was the choice of uniform array switching patterns that causes this limit in \gls{tdm} channel sounding \cite{yin2003doppler}. A non-uniform array switching pattern can potentially significantly extend the estimation range of Doppler shifts and eliminate the ambiguities. They studied the problem in the context of the ISI-SAGE algorithm \cite{fleury2003high}. However their analysis not only assumes that all antenna elements are isotropic radiators, but also requires the knowledge of the phase centers of all antennas. Both assumptions are difficult to fulfill for realistic arrays used in a channel sounder, given the mutual coupling between antennas and the presence of a metallic support frame. Pedersen \etal proposed to use the so-called \gls{nsl} of the objective function as the metric to identify switching patterns \cite{pedersen2004joint}. They further derived a necessary and sufficient condition of the array switching sequence that leads to ambiguities \cite{pedersen2008optimization}, but the method to design a good switching sequence is not clear. To our best knowledge, no methods have been proposed to find optimized switching patterns for realistic arrays.

Our work adopts an algebraic model that uses decompositions through \glspl{eadf} \cite{belloni2007doa}, which provides a reliable and elegant approach for signal processing with real-world arrays. Therefore our analysis no longer requires the isotropic radiation pattern or prior knowledge about the antenna phase centers. Based on the Type \RNum{1} ambiguity function for an arbitrary array \cite{eric1998ambiguity}, we propose a spatio-temporal ambiguity function and investigate its properties and impact on the estimation of directions and Doppler shifts of \glspl{mpc}. Inspired by Ref. \cite{chen2008mimo}, we model the array pattern design problem as an optimization problem and propose an algorithm based on \gls{sa} to search for an acceptable solution. The results are validated with Monte-Carlo simulations when the final switching sequence is incorporated into a RiMAX-based \gls{hrpe} algorithm.

The remainder of the paper is organized as follows. Section \RNum{2} introduces the signal data model in \gls{tdm} channel sounding and the spatio-temporal ambiguity function. In Section \RNum{3} we present the formulation of the optimization problem and its solution based on the \gls{sa} algorithm. In Section \RNum{4} we demonstrate the performance of the optimal switching sequence with a corresponding \gls{hrpe} algorithm, and compare \glspl{rmse} of key parameters with the squared root of \gls{crlb}. In Section \RNum{5} we draw the conclusions.

\section{Signal Model and ambiguity function}
\subsection{Signal Data Model}
This work mainly studies the antenna switching sequence in the \gls{tdm} channel sounding problem. We consider $T$ \gls{mimo} measurement snapshots in one observation window, each with $M_f$ frequency points, $M_R$ receive antennas, and $M_T$ transmit antennas. The adjacent \gls{mimo} snapshots are separated by $T_0$. We assume that all scatterers are placed in the far field of both \gls{tx} and \gls{rx} arrays, which also implies that \glspl{mpc} are modeled as plane waves. Besides \gls{tx} and \gls{rx} arrays are vertically polarized by assumption and have frequency-independent responses within the operating bandwidth. Such $T$ \gls{mimo} snapshots can span larger than the coherence time of the channel, but we assume that the large-scale parameters of the channel, such as path delay, \gls{doa}, \gls{dod} and Doppler shift, remain constant during this period.\footnote{This does not indicate the channel is assumed to be static, the complex path weight of SP have a phase rotation due to the presence of Doppler shift.}     

A \textit{vectorized} data model for the observation of $T$ \gls{mimo} snapshots is given in Eq. (\ref{Eq:y_VecModel}). It includes contributions from deterministic \glspl{sp} $\B{s}(\BS\theta_{sp})$,  \gls{dmc} $\B{n}_{dmc}$ and measurement noise $\B{n}_0$.
\begin{equation}
 \B{y} = \B{s}(\BS\theta_{sp}) + \B{n}_{dmc} + \B{n}_{0}  \label{Eq:y_VecModel}
\end{equation}
The data model of a observation vector for a total number of $P$ \glspl{sp} is determined by
\begin{equation}
 s(\BS\theta_\tx{sp}) = \B{B}(\BS\mu) \cdot \BS\gamma_{vv}, \label{Eq:s_theta},
\end{equation}
where $\BS\gamma_{vv}$ is a vector of the complex path gains, $\BS\mu$ is the structural parameter vector, and $\B{B}(\BS\mu)$ is the basis matrix. The detailed structure of $\B{B}(\BS\mu)$ when a uniform switching pattern is used can be found in \cite[Eq. (20)]{WangHRPE2017}. $\BS\theta_{sp}$ is the entire parameter vector for \glspl{sp}, thus includes both $\BS\mu$ and $\BS\gamma_{vv}$.

We would like to incorporate the choice of non-uniform switching patterns into this signal data model. To faciliate the implementation of the associated \gls{hrpe} algorithm, we only allow the \gls{tx} array to implement a cycle-dependent switching pattern, while the \gls{rx} array's switching pattern remains uniform. As a result the basis matrix can be expressed as
\begin{align}
  \B{B}(\BS{\mu}) &= \tilde{\B{B}}_{TV,T} \diamond \tilde{\B{B}}_{RV} \diamond \B{B}_f   \\
  &=\begin{bmatrix}
  \tilde{\B{B}}_{TV}^1  & \cdots  &\tilde{\B{B}}_{TV}^T 
 \end{bmatrix}^T  \diamond \tilde{\B{B}}_{RV} \diamond \B{B}_f,
 \label{Eq:BasisMatrix_Simplify}
\end{align}
where $\B{B}_{TV}^j \in \mathbb{C}^{M_TT\times P}$ with $j=1,2,\ldots,T$ represents the spatio-temporal response of \gls{tx} array at the $j$-th MIMO snapshot, and $\tilde{\B{B}}_{RV} \in \mathbb{C}^{M_R \times P}$ is for the \gls{rx} array with uniform switching, and $\B{B}_f \in \mathbb{C}^{M_f \times P}$ is the basis matrix that captures the frequency response due to path delay. Compared with the basis matrix in \cite[Eq. (20)]{WangHRPE2017}, we replace $\B{B}_t \diamond \tilde{\B{B}}_{TV}$ with $\tilde{\B{B}}_{TV,T}$ due to the scrambled switching sequence allowed at the \gls{tx} array.

Since our work focuses on channel sounding in a fast time-varying channel, the phase variation within one \gls{mimo} snapshot is no longer negligible, and the new \gls{tx} or \gls{rx} basis matrices for the $t$-th \gls{mimo} snapshot become weighted versions of the static array basis matrices. The exact connections are given by
\begin{align}
 \tilde{\B{B}}_{TV}^t &= \B{B}_{TV} \odot \B{A}_T^t \\
 \tilde{\B{B}}_{RV} &= \B{B}_{RV} \odot \B{A}_R, 
\end{align}
where $\B{A}_R$ and $\B{A}_T^t$ are weighting matrices that capture the phase change due to effects of Doppler and switching schemes. $\B{A}_T^t$ depends on the \gls{mimo} snapshot index $t$, because \gls{tx} implements a cycle-dependent switching pattern. Let us use a $M_T \times T$ matrix $\BS\eta_T$ to represent the \gls{tx} switching pattern, the elements in the phase weighting matrices $\B{A}_T^t$ and $\B{A}_R$ are given by

\begin{align}
 [\B{A}_T^t]_{m_T,p} &= e^{j2\pi\nu_p [\BS\eta_T]_{m_T,t}} \\
 [\B{A}_R]_{m_R,p} &= e^{j2\pi\nu_p m_Rt_0}.
\end{align}
If we maintain the duration between two adjacent switching events for both \gls{tx} and \gls{rx} array, and denote them as $t_1$ and $t_0$ respectively, we have
\begin{equation}
  [\BS\eta_T]_{m_T,t} = (t-1)M_Tt_1 + ([\B{S}_T]_{m_T,t} - 1)t_1
\end{equation}
$[\B{S}_T]_{m_T,t}$ takes an integer value between $1$ and $M_T$ and represents the scheduled switching index of the $m_T$th \gls{tx} antenna for the $t$th \gls{mimo} snapshot. For example with the uniform switching pattern, we have $[\B{S}_T]_{m_T,t} = m_T$, $\forall t=1,2,...,T$.

\subsection{Spatio-temporal Ambiguity Function}
The Type \RNum{1} ambiguity function for an antenna array can reflect its ability to differentiate signals in the angular domain \cite{eric1998ambiguity}. Generalizing the definition to include the full parameter vectors employed in our model, we have
 \begin{equation}
    X_{tot}(\BS{\mu}_1,\BS{\mu}_2) =  \frac{ \B{b}(\BS{\mu}_1,\BS{\eta}_T)^\dagger\B{b}(\BS{\mu}_2,\BS{\eta}_T)}{\lVert\B{b}(\BS{\mu}_1,\BS{\eta}_T) \rVert \cdot \lVert\B{b}(\BS{\mu}_2,\BS{\eta}_T) \rVert}, \label{Eq:Amb_func}
 \end{equation}
where $\BS\mu_1$ or $\BS\mu_2$ stands for the structural parameter vector of \textit{one} \gls{sp}. $\B{b}(\BS{\mu}_1,\BS{\eta}_T)$ can be treated as one particular column of $\B{B}(\BS\mu)$.

This multi-dimensional ambiguity function is also closely related to the ambiguity function well studied in \gls{mimo} radar. The \gls{mimo} radar ambiguity function in Ref. \cite{chen2008mimo} allows the \gls{tx} to send different waveforms on different antennas, while our problem considers a single sounding waveform for all \gls{tx} antennas. The Doppler-(bi)direction ambiguity function used in Ref. \cite{pedersen2008optimization} is also quite similar to ours, except that we work with the channel transfer function instead of the received waveform. Moreover they assume that the array is composed of identical elements and suffers no mutual coupling effects, our ambiguity function is more general in the sense that it handles various array structures. 

Additionally instead of directly evaluating the ambiguity function in Eq. (\ref{Eq:Amb_func}), we can find an upper bound for its amplitude, which merely depends on the azimuth \gls{dod}, the Doppler shift, and the \gls{tx} switching pattern. Our work considers a switched array setup in a fast time-varying channel, and the measurement spans both spatial and temporal domain, so it is natural to focus on two dimensions, namely azimuth \gls{dod} and Doppler. We call the simplified 2D ambiguity function the spatio-temporal ambiguity function, and its relation with the generalized ambiguity function of Eq. (\ref{Eq:Amb_func}) is given as
\begin{equation}
  \vert X_{tot}(\BS{\mu},\BS{\mu}^\prime) \vert \le \vert X_T(\varphi_T.\varphi_T^\prime,\nu,\nu^\prime)\vert. \label{Eq:2ndUpperBound}
\end{equation}
Detailed derivations about this inequality and associated properties of this spatio-temporal ambiguity function are included in Ref. \cite{Wang2018Sw}. Furthermore we can show that this upper bound only depends on the difference of two Doppler shifts, $\ie$ $\Delta\nu = \nu^\prime - \nu$, which is given by
\begin{equation}
  X_T(\varphi_T,\varphi_T^\prime,\nu,\nu^\prime) = \frac{\tilde{\B{b}}_{TV,T}^\dagger(\varphi_T,\nu) \tilde{\B{b}}_{TV,T}(\varphi_T^\prime,\nu^\prime)}{\lVert \tilde{\B{b}}_{TV,T}^\dagger(\varphi_T,\nu) \rVert \lVert \tilde{\B{b}}_{TV,T}(\varphi_T^\prime,\nu^\prime) \rVert}. \label{Eq:X_T_analytic}
\end{equation}
Its nominator can be written as 
\begin{align}
 \sum_{t = 1}^T &\tilde{\B{b}}_{TV,T}^t (\varphi_T,\nu) ^\dagger \tilde{\B{b}}_{TV,T}(\varphi_T^\prime,\nu^\prime) \notag \\
  &= \Big[\B{b}_{TV}(\varphi_T) \odot \B{b}_{TV}^*(\varphi_T^\prime)\Big]^\dagger \sum_{t=1}^T e^{j2\pi(\nu^\prime - \nu)\BS\eta_T^t},
\end{align}
and it only depends on $\Delta\nu$. 
The denominator of Eq. (\ref{Eq:X_T_analytic}) can be expressed by
\begin{equation}
  \lVert \tilde{\B{b}}_{TV,T}(\varphi_T,\nu) \rVert = \sqrt{T} \cdot \lVert \B{b}_{TV}(\varphi_T) \rVert,
\end{equation}
and it is independent of $\nu$. 

\begin{figure}[!t]
 \centering
 \includegraphics[width = 3.5in]{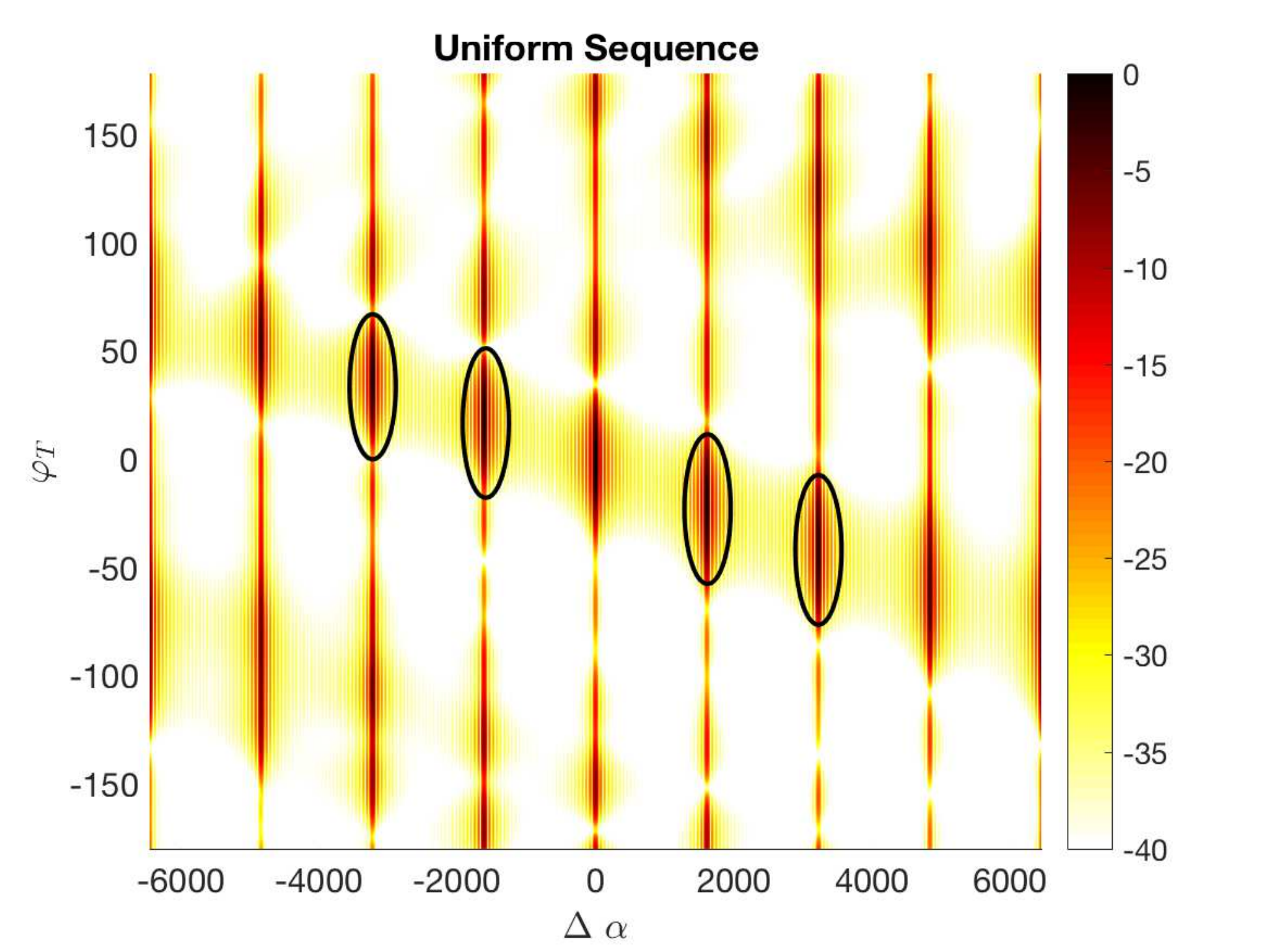}
 \caption{Amplitude of Ambiguity function (dB) with Azimuth \gls{dod} and Doppler shift, under uniform switching schemes at both Tx and Rx}
 \label{fig:ChiAmp_unif_exp}
\end{figure}

It is well known that the Doppler shifts and the impinging directions of the plane waves may contribute to phase changes at the output of the array. The Doppler shift leads to a phase rotation at the same antenna when it senses at different time instants, while the propagation direction of the plane wave also contributes to a phase change between two antennas. The \textit{periodic} structure of the uniform switching sequence leads to ambiguities in the joint estimation of Doppler and propagation direction. It is because the estimator may find more than one plausible combination of Doppler and angle that can produce the phase changes over different antennas and time instants. For example Fig. \ref{fig:ChiAmp_unif_exp} plots the amplitude upper bound of the ambiguity function given in Eq. (\ref{Eq:2ndUpperBound}), when the \gls{tx} uses the uniform switching pattern. We observe multiple peaks in addition to the central peak located at $(0,0)$. It also shows that the non-ambiguous estimation range of Doppler is $[-1/2T_0,1/2T_0)$, when both \gls{tx} and \gls{rx} implement uniform switching patterns and $\varphi_T^\prime = 0$. The value of $T_0$ in this example is $\SI{620}{\mu s}$ and is based on the transmitted signal in Ref. \cite{wang2017real}. 

\section{Algorithm}

To solve the sequence design problem we need a reasonable metric to evaluate the performance of a particular switching sequence. More importantly we intend to apply the results to wireless channel sounding application, therefore the problem formulation and performance metric should be suitable for real-world arrays.

\subsection{Problem Formulation}
The intuitive objective of our array switching design problem is to find schemes that effectively suppress the sidelobes of the spatial-temporal ambiguity function shown in Fig. \ref{fig:ChiAmp_unif_exp}, hence increases the estimation range of Doppler shift. The authors in Refs. \cite{pedersen2008optimization} and \cite{chen2008mimo} prove that their ambiguity functions have a constant energy, so a preferable scheme should spread the volume under the high sidelobes evenly elsewhere. However the proof again uses the idealized assumption about antenna arrays, thus it cannot be applied directly in our case. Here we introduce the function $f_p(\BS\eta_T)$, which is given by 
\begin{align}
 f_p(\BS\eta_T) &= \underset{D}{\iiint} \Big|X_T(\varphi_T,\varphi_T^\prime,\Delta\nu)\Big|^p\; \mathrm{d}\varphi_T  \, \mathrm{d}\varphi_T^\prime  \, \mathrm{d}\Delta\nu, \\
 D &= \{(\varphi_T,\varphi_T^\prime,\Delta\nu) \vert \varphi_T,\varphi_T^\prime \in (\pi,\pi] \,\& \,\Delta\nu \in [0,\nu_\tx{up}] \}\notag
\end{align}
where $D$ is the integration intervals, and $\nu_\tx{up}$ represents the target maximal Doppler shift within which we want to avoid the ambiguity. This value should equal $M_T/2T_0$, because the periodic sequence is only $T_0/M_T$ long. We have conducted numerical simulations based on our proposed ambiguity function in Eq. (\ref{Eq:2ndUpperBound}), with the calibrated response of the \gls{tx} array used in a \gls{v2v} \gls{mimo} channel sounder \cite{wang2017real}. Results have shown that its energy, i.e. $f_2(\BS\eta_T)$, is almost constant regardless of the choices of different $\BS\eta_T$. 
As a result we can use $f_p(\BS\eta_T)$ with a higher value of $p$ as the cost function to penalize the \gls{tx} switching schemes that lead to high sidelobes. As a summary our optimization problem is given by
\begin{align}
 \underset{\B{S}_T \in \mathcal{C}}{\tx{min}} \; &f_p(\BS\eta_T)\\
 \text{s.t.}\;  &[\BS\eta_T]_{m_T,t} = ([\B{S}_T]_{m_T,t} - 1)t_1 + (t-1)M_Tt_1, \notag
\end{align}
where the elements in the set $\mathcal{C}$ are integer matrices with a dimension of $M_T \times T$, and every column of $\B{S}_T$ is a permutation of the vector $[1,2,\ldots,M_T]^T$. 

\subsection{Solution and Results}
Because $\B{S}_T$ takes on discrete values in the feasible set, the simulated annealing algorithm is known to solve this type of problem \cite{kirkpatrick1983optimization}. The pseudocode of our proposed algorithm is given here in Alg. \ref{alg:anneal}.
\begin{algorithm}[h]
 \caption{The annealing algorithm to solve our RSAA design problem}
 \begin{algorithmic}[1]
   \STATE Initialize $\BS\eta_T$, the temperature $T = T_0$, and $\alpha = \alpha_0$;
   \WHILE{ $k\le k_\tx{max}$ or $f_p(\BS\eta_T) > \epsilon_{th}$}
      \STATE $\BS\eta_T^\prime =$ neighbor($\BS\eta_T$);
      \IF{$\tx{exp}\big[ (f_p(\BS\eta_T) - f_p(\BS\eta_T^\prime))/T \big] > \tx{random}(0,1)$}
         \STATE $\BS\eta_T = \BS\eta_T^\prime$
      \ENDIF
      \STATE $T = \alpha T$
   \ENDWHILE
 \end{algorithmic}
 \label{alg:anneal}
\end{algorithm}
The key parameters related with this algorithm are $p=6$, the initial temperature is $T_0=100$, the cooling rate is $\alpha_0 = 0.97$ and the $k_\tx{max}=500$.

The transition probability from switching scheme (state) $\BS\eta_T$ to $\BS\eta_T^\prime$ is chosen as
\begin{align}
 &P_r(\BS\eta_T,\BS\eta_T^\prime) = \notag\\
 &\begin{cases}
  \frac{1}{A} \tx{min}\big(1,\tx{exp}(\frac{f_p(\BS\eta_T) - f_p(\BS\eta_T^\prime)}{T}) \big), &\tx{if } \BS\eta_T^\prime \sim \BS\eta_T \\
    1 - \frac{1}{A} \underset{\BS\eta_T^{\prime\prime} \sim \BS\eta_T}{\sum} \tx{min} \big(1,\tx{exp}(\frac{f_p(\BS\eta_T) - f_p(\BS\eta_T^{\prime\prime})}{T}) \big), &\tx{if }\BS\eta_T^\prime = \BS\eta_T \\
  0, & \tx{otherwise}
 \end{cases}
\end{align}
where $\BS\eta_T^\prime \sim \BS\eta_T$ denotes that $\BS\eta_T^\prime$ and $\BS\eta_T$ only differ by swapping two elements in the same column. $A$ is the cardinality of the set $\{\BS\eta_T^\prime\vert \BS\eta_T^\prime \sim \BS\eta_T\}$, i.e. $A = T\binom{M_T}{2}$.

\begin{figure}[!t]
 \centering
 \includegraphics[width = 3.5in]{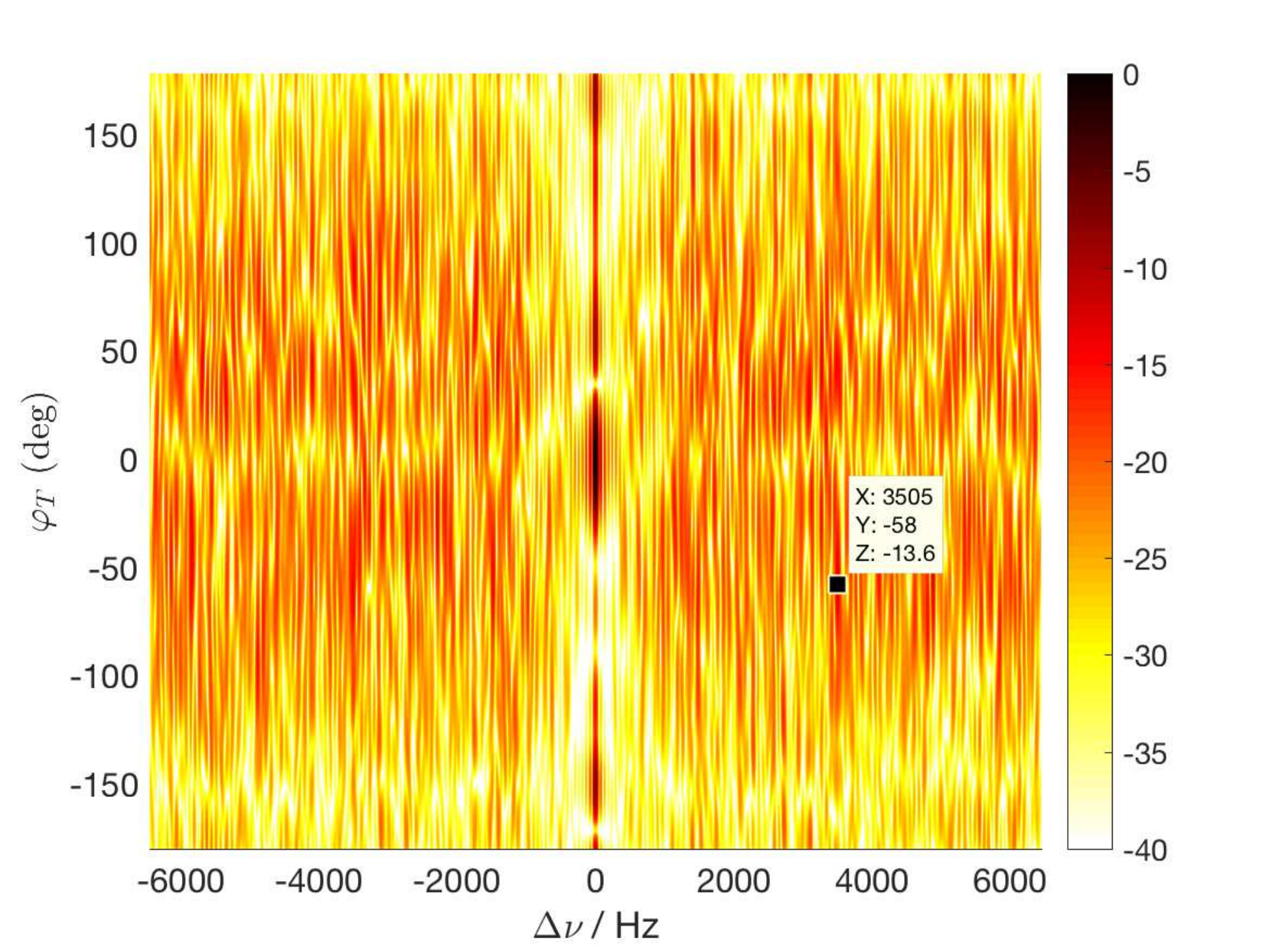}
 \caption{Amplitude of Ambiguity function with Azimuth \gls{dod} and Doppler shift, under uniform Rx switching and scrambled Tx switching scheme}
 \label{fig:ChiAmp_rand_exp}
\end{figure}

\begin{figure}[!t]
 \centering
 \includegraphics[clip = true, width = 0.95\columnwidth]{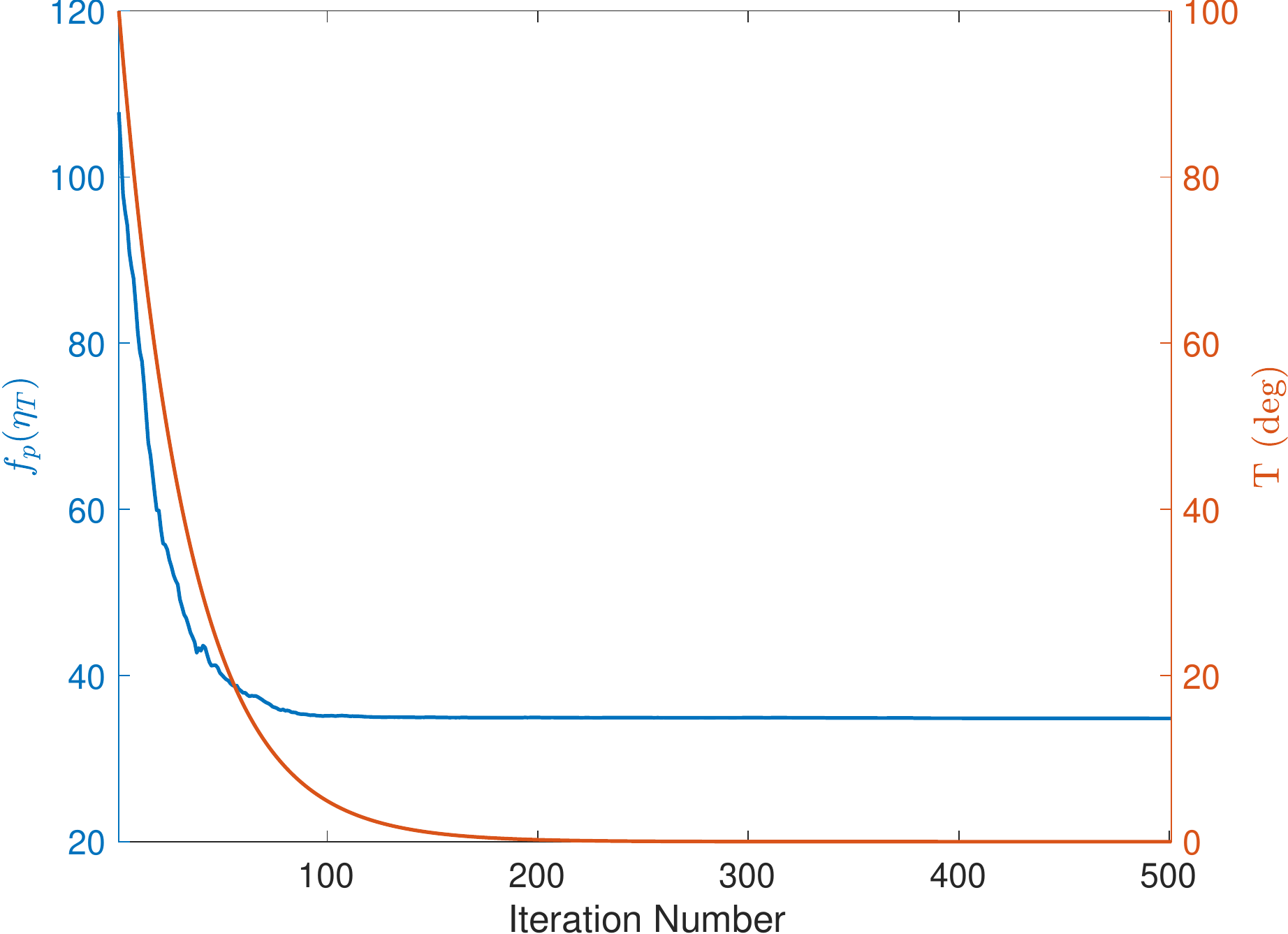}
 \caption{The evolution of $f_6(\BS\eta_T)$ and temperature during the annealing algorithm in Algorithm \ref{alg:anneal}}
 \label{fig:f_p_evolve}
\end{figure}

Fig. \ref{fig:f_p_evolve} provides the values of the objective function with the iteration number and decreased temperature in the \gls{sa} algorithm. We present the amplitude of the 2D ambiguity function (the upper bound given in Eq. (\ref{Eq:2ndUpperBound})) with the final switching sequence in Fig. \ref{fig:ChiAmp_rand_exp}, where the high sidelobes clearly disappear in constrast with Fig. \ref{fig:ChiAmp_unif_exp}. Another useful metric is the \gls{nsl} used in Ref. \cite{pedersen2004joint} to measure the quality of the switching sequence. The \gls{nsl} here is \SI{-13.60}{dB}, where the lowest value provided in Ref. \cite{pedersen2004joint} is \SI{-11.06}{dB}. 

\section{Numerical Simulations}
In this section we use Monte Carlo simulations to validate the choice of the switching sequence by investigating the \glspl{rmse} of some key parameters, such as azimuth \gls{dod} and Doppler shift. We also manage to incorporate $\BS\eta_{T,f}$ into an \gls{hrpe} algorithm based on the framework of RiMAX, for more details see Ref. \cite{Wang2018Sw}. As a comparison we also investigate the case when the same channel is sampled with uniformly switched arrays and evaluated with the corresponding \gls{hrpe} algorithm in Ref. \cite{WangHRPE2017}.

\begin{figure}[!t]
  \centering
   \includegraphics[width = 0.8\columnwidth,clip=true]{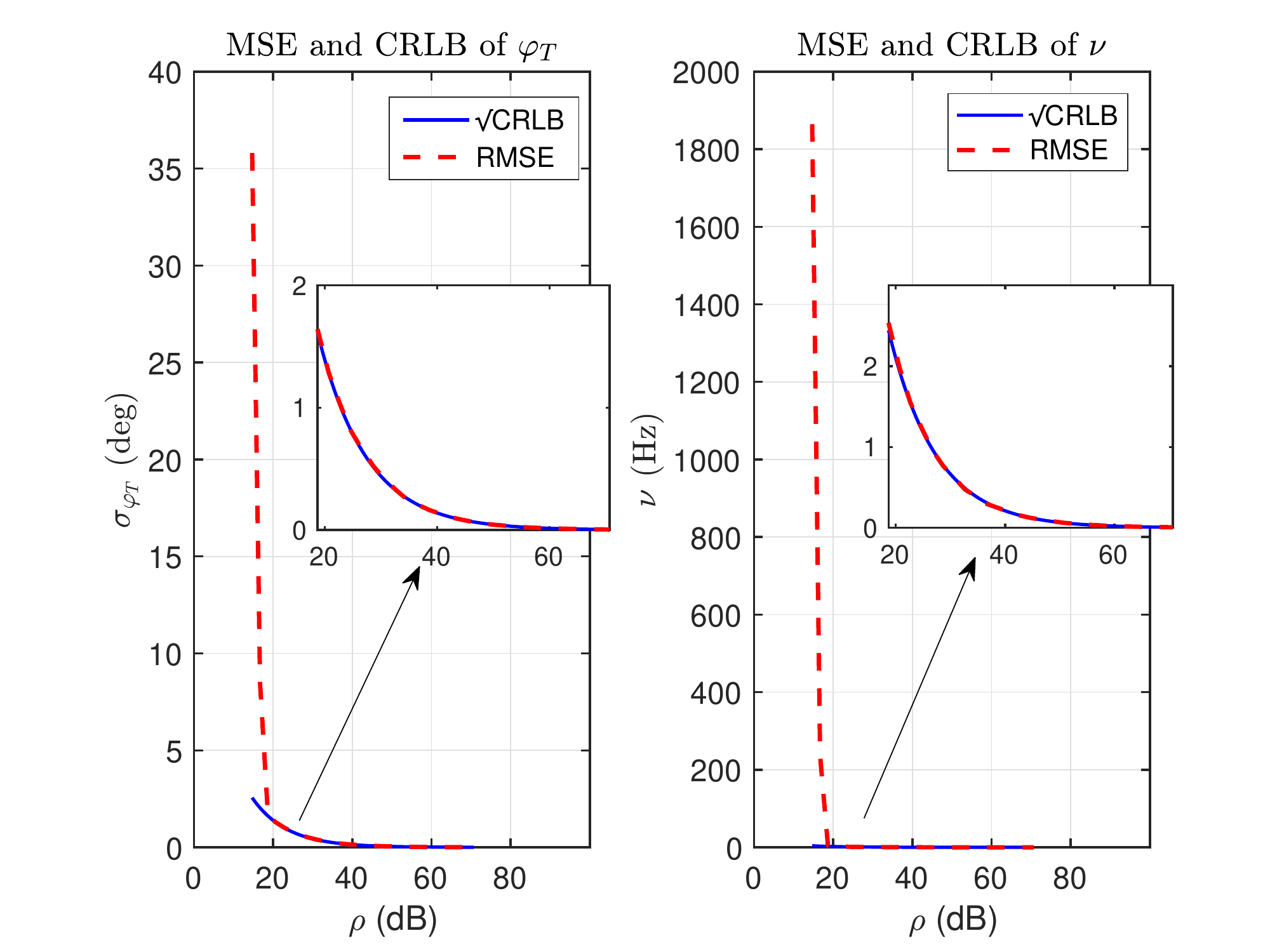}
  \caption{Comparison about RMSE of $\varphi_T$ and $\nu$ based on Monte-Carlo simulations with the theoretical \gls{crlb} from Eq. (\ref{Eq:crlb_theta_s}) for snapshot 1 in Tab. \ref{Tab:one-path_Ch_param} for our scrambled switching pattern.}
  \label{fig:crlb_case1_rs}
\end{figure}

\begin{figure}[!t]
  \centering
  \includegraphics[width = 0.8\columnwidth,clip=true]{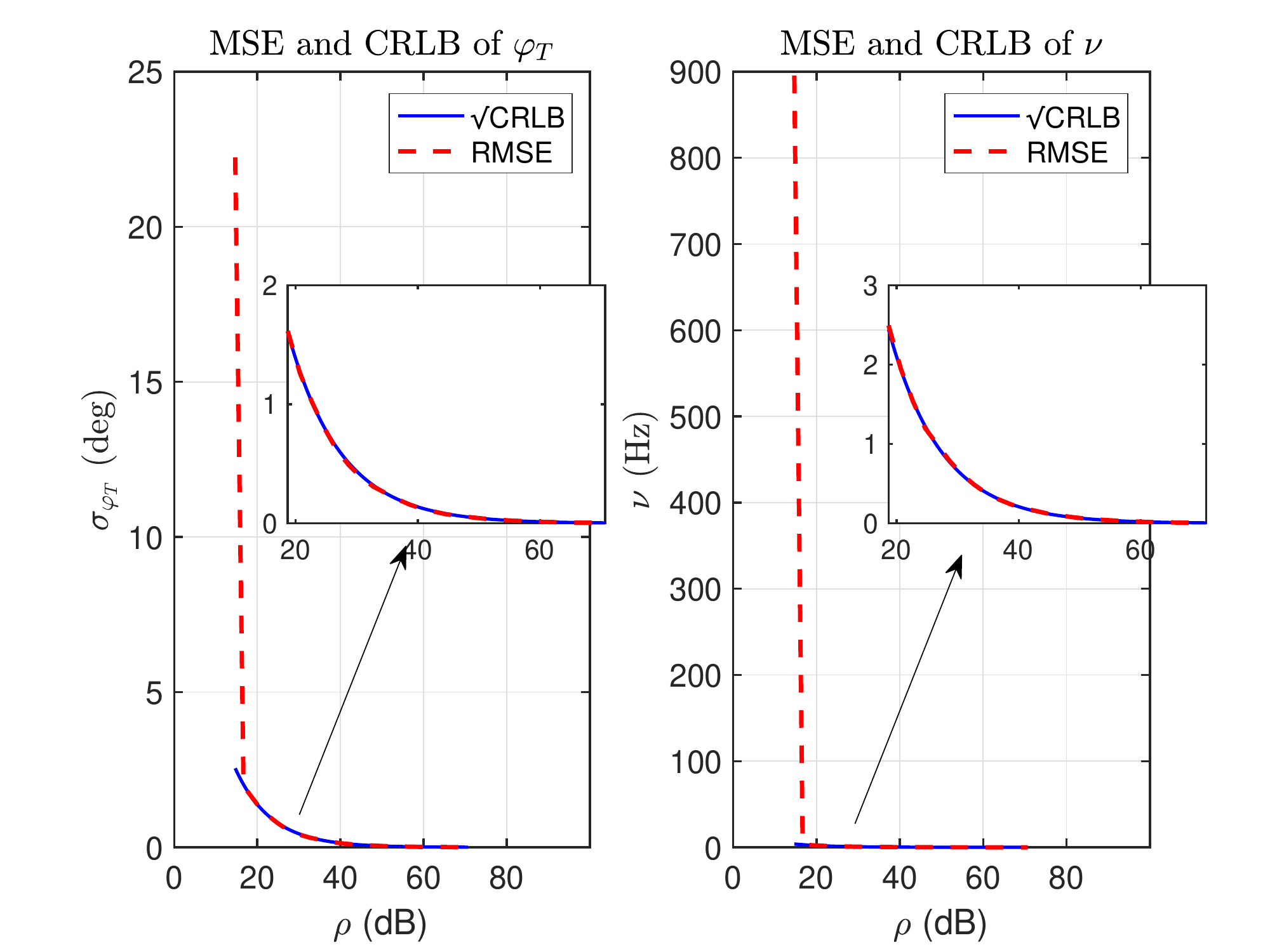}
  \caption{Comparison about RMSE of $\varphi_T$ and $\nu$ based on Monte-Carlo simulations with the theoretical \gls{crlb} from Eq. (\ref{Eq:crlb_theta_s}) for snapshot 2 in Tab. \ref{Tab:one-path_Ch_param} for our scrambled switching pattern.}
  \label{fig:crlb_case2_rs}
\end{figure}

\begin{figure}[!t]
  \centering
   \includegraphics[width = 0.8\columnwidth,clip=true]{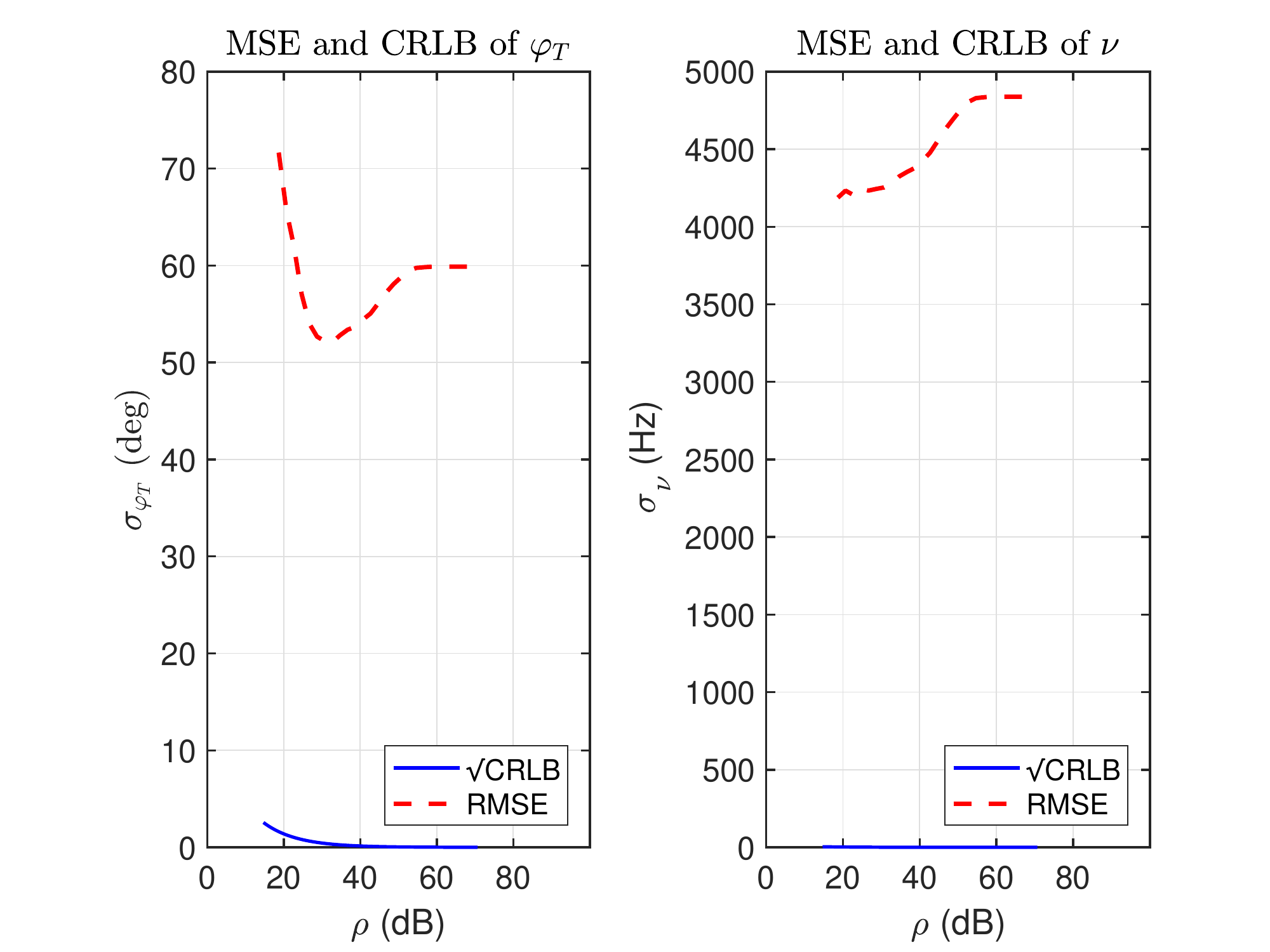}
  \caption{Comparison about RMSE of $\varphi_T$ and $\nu$ based on Monte-Carlo simulations with the theoretical \gls{crlb} from Eq. (\ref{Eq:crlb_theta_s}) for snapshot 1 in Tab. \ref{Tab:one-path_Ch_param} for uniform switching pattern.}
  \label{fig:crlb_case1_unif}
\end{figure}

\begin{figure}[!t]
  \centering
  \includegraphics[width = 0.8\columnwidth,clip=true]{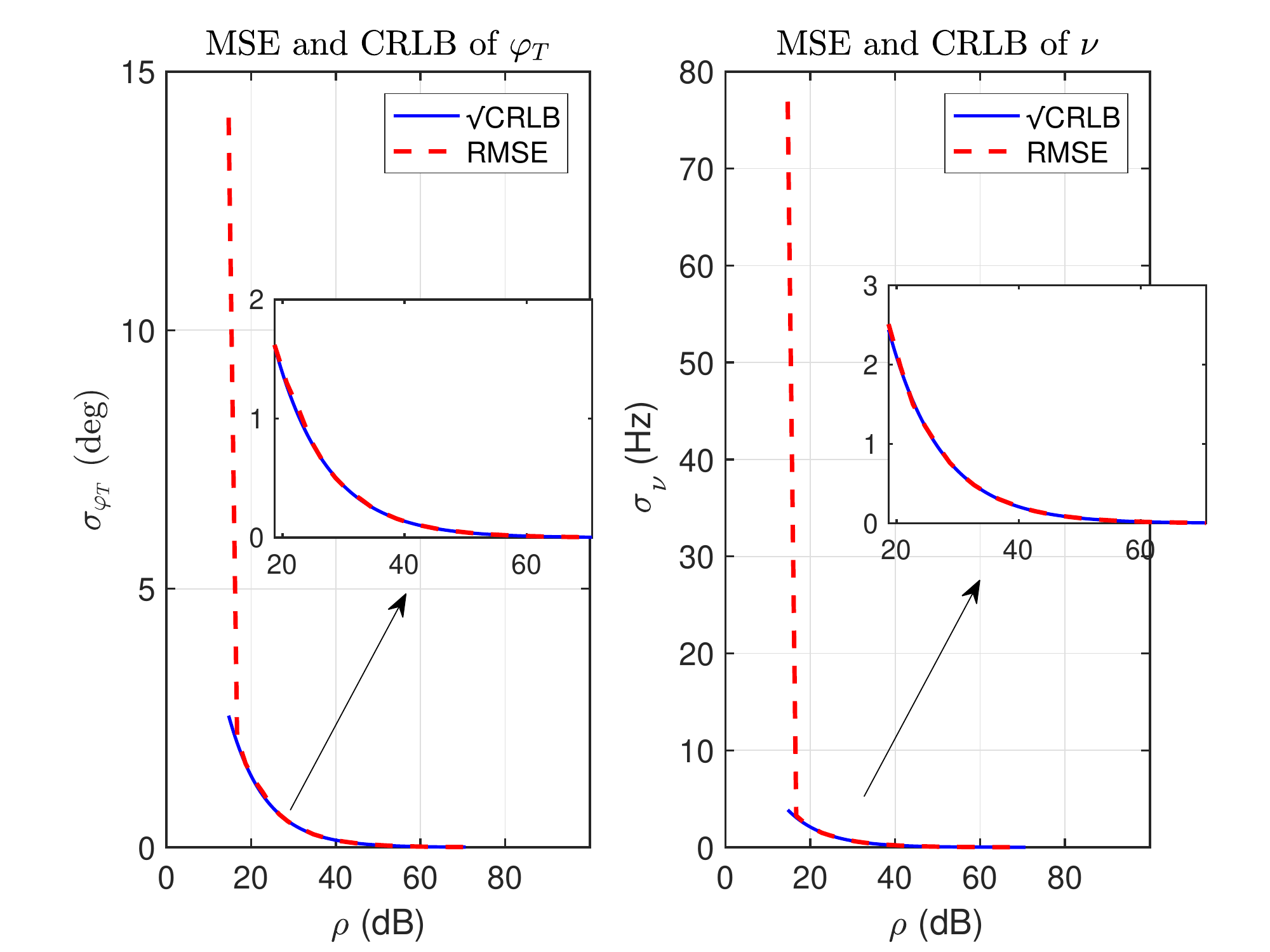}
  \caption{Comparison about RMSE of $\varphi_T$ and $\nu$ based on Monte-Carlo simulations with the theoretical \gls{crlb} from Eq. (\ref{Eq:crlb_theta_s}) for snapshot 2 in Tab. \ref{Tab:one-path_Ch_param} for uniform switching pattern.}
  \label{fig:crlb_case2_unif}
\end{figure}

For clarity of definitions, we analyze an environment with a single SP, whose parameters are listed in Tab. \ref{Tab:one-path_Ch_param}. To cover several cases of interest, $\vert\nu\vert$ is larger than $1/2T_0\approx\SI{806}{Hz}$ in snapshot 1 and smaller than $1/2T_0$ in snapshot 2. We compare the simulated \glspl{rmse} with the squared root of \gls{crlb} as a function of \gls{snr} $\rho$ for two switching sequences, which are the uniform sequence $\BS{\eta}_{T,u}$ and our optimized scrambled \gls{tx} sequence $\BS{\eta}_{T,f}$. We simulate 1000 realizations at each \gls{snr} value. The theoretical \gls{crlb} can be determined based on \gls{fim} and given by
\begin{equation}
  \sigma_{\BS\theta_s}^2 \succeq \tx{diag}(\mathcal{J}^\tx{-1}(\BS\theta_s)),
  \label{Eq:crlb_theta_s}
\end{equation}
where $\succeq$ is the generalized inequality for vectors. Figs. \ref{fig:crlb_case1_rs} and \ref{fig:crlb_case2_rs} provide such a comparison for $\BS{\eta}_{T,f}$, which demonstrates its good performance in both channels with high or low Doppler. On the other hand, Fig. \ref{fig:crlb_case1_unif} shows the poor estimation accuracy in the high Doppler case for the case of uniform switching pattern $\BS{\eta}_{T,u}$, although the \gls{mse} can achieve \gls{crlb} in the low Doppler scenario as expected in Fig. \ref{fig:crlb_case2_unif}. 

\begin{figure}[!t]
  \centering
  \includegraphics[width = 0.9\columnwidth,clip=true]{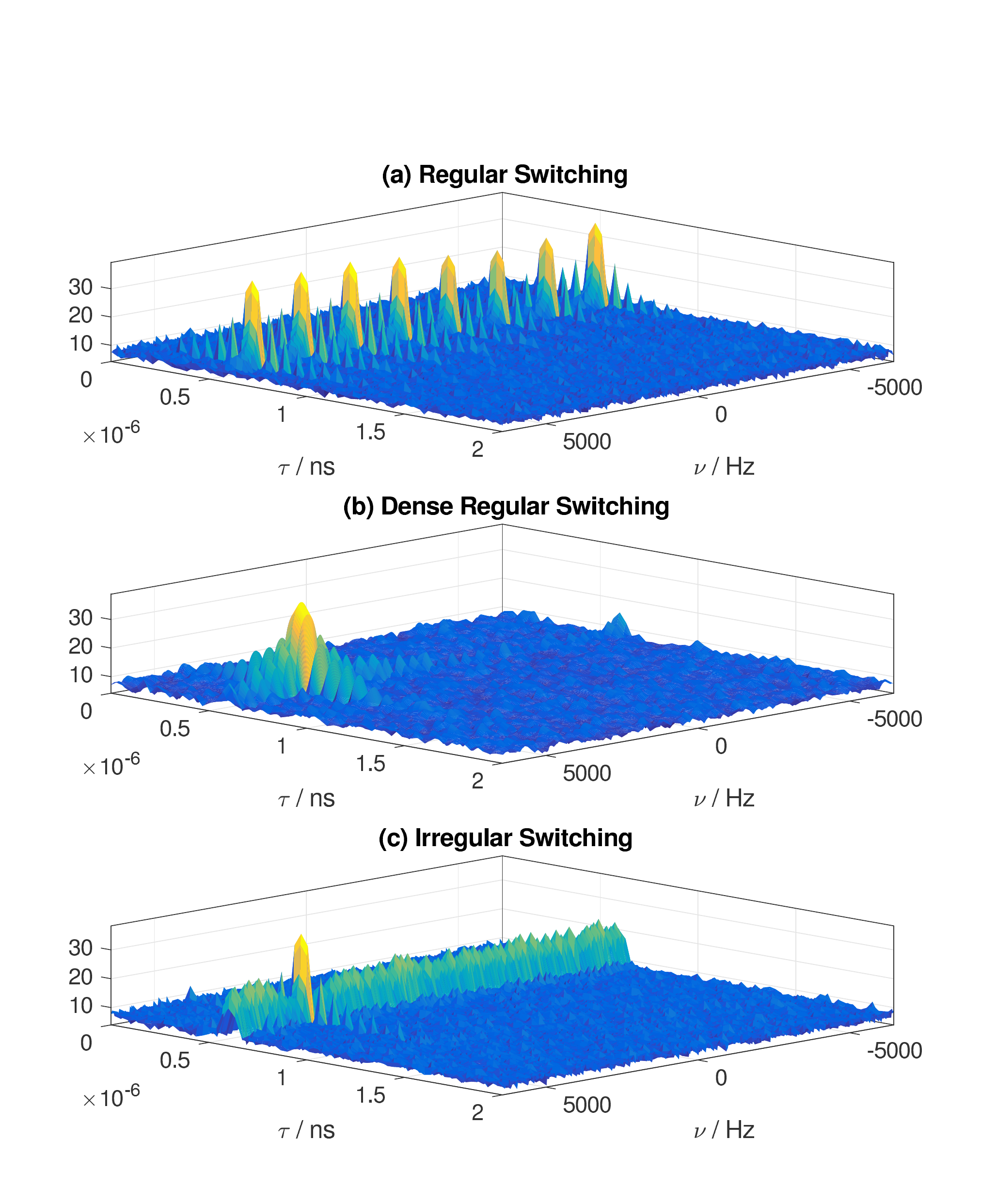}
  \caption{Delay-Doppler spectrum of snapshot 1 with large Doppler under different TX switching sequences, (a) $\BS\eta_{T,u}$ (b) ``dense'' uniform sequence (c) $\BS\eta_{T,f}$}
  \label{fig:Delay_Doppler_Spec}
\end{figure}

We also shown in Fig. \ref{fig:Delay_Doppler_Spec} the delay-Doppler spectrum of snapshot 1 with three different \gls{tx} switching sequences, which are $\BS\eta_{T,u}$, $\BS\eta_{T,d}$ (known as the ``dense'' uniform sequence), and $\BS\eta_{T,f}$. As a result, the spectrum in Fig. \ref{fig:Delay_Doppler_Spec}(a) displays multiple peaks in the same delay bin but at different Doppler shifts, while Fig. \ref{fig:Delay_Doppler_Spec}(c) shows that $\BS\eta_{T,f}$ successfully eliminates all the peaks except for one at the desired location. Notice that $\BS\eta_{T,f}$ also helps distribute the power under those unwanted peaks equally across Doppler. Fig. \ref{fig:Delay_Doppler_Spec}(b) shows that with $\BS\eta_{T,d}$, we can also eliminate the repeated main peaks (and achieve slightly lower sidelobe energy); however, at the price of the separation time between adjacent \gls{mimo} snapshots is reduced to 1/8 of $T_0$ in $\BS\eta_{T,u}$, which would imply that the number of antenna elements would have to be reduced such that $M_T M_R$ decreases by a factor of 8.

\begin{table}[!t]
  \centering
  \caption{Path parameters of one-path scenario, which serves to verify the RMSEs of $\varphi_T$ and $\nu$}
  \label{Tab:one-path_Ch_param}
  \begin{tabular}{l|c|c|c|c}
    \toprule
    Snapshot & $\tau$ (ns) & $\varphi_T$ (deg) & $\varphi_R$ (deg) & $\nu$ (Hz) \\
    \midrule
    1        &  601.1      & 11.5 & 59.6 & 4032.3  \\
    2        &   1117.3    & 21.3 & 160.0 & 80.6   \\
    \bottomrule 
  \end{tabular}
\end{table}

Besides we simulate a two-path channel, which is the simplest version of the multipath channel. Tab. \ref{Tab:case5_est} provides a comparison between the true and estimated parameters where we apply $\BS{\eta}_{T,f}$ and the newly developed \gls{hrpe} algorithm. Although both paths' Doppler shifts are larger than $1/2T_0$, the simulation suggests that the estimated parameters are pleasingly close to the true values.

\begin{table}[!t]
  \tiny
  \centering
 \caption{Parameter estimation for a two-path channel with $\BS{\eta}_{T,f}$, true/estimate}
 \label{Tab:case5_est}
  \begin{tabular}{c|c|c|c|c|c}
    \toprule
    Path ID & $\tau$ (ns) & $\varphi_T$ (deg) & $\varphi_R$ (deg) & $\nu$ (Hz) & $\vert\gamma\vert^2$ (dB) \\
    \midrule
    1& 646.2/646.2 & 67.81/67.79 & -59.33/-59.33 & 3225.8/3225.8 & -13.13/-13.13\\
    2& 1203.7/1203.7 & -60.15/-60.15 & -123.79/-123.78 & 3217.7/3217.8 & -18.82/-18.82\\
    \bottomrule
  \end{tabular}  
\end{table}

\section{Conclusion}
In this paper we revisit the array switching design problem and focus on its adaptability to realistic arrays deployed in channel sounders for fast time-varying channels. We formulate the switching design problem as an optimization problem to suppress the sidelobes of the spatio-temporal ambiguity function, which is dependent on array switching patterns. We also show through Monte-Carlo simulations that a RiMAX-based \gls{hrpe} algorithm where the new sequence is incorporated can produce accurate Doppler and angular estimates in both high and low Doppler scenarios. In the future we will further validate this method and implement related results in applications such as \gls{v2v} or mmWave \gls{mimo} channel measurements. 

\section*{Acknowledgment}
This work was supported by grants from Caltrans, the National Institute of Standards and Technology and the National Science Foundation.



%

\bibliographystyle{IEEEtran}
\bibliography{SwSeq_ref}

\end{document}